\documentclass[fleqn,usenatbib]{mnras}
\usepackage{amsmath}
\usepackage{amssymb}
\usepackage{amsfonts}
\usepackage{subcaption}
\usepackage{nicefrac}
\usepackage[T1]{fontenc}
\usepackage{graphicx}
   
\renewcommand{\selectlanguage}[1]{}
\global\long\def\rmd{\mathrm{d}}%
\global\long\def\f#1#2{\frac{#1}{#2}}%
\global\long\def\d{\delta}%
\global\long\def\D{\Delta}%
\newcommand{\FEWBODY}{\texttt{FEWBODY}}
\global\long\def\s#1{\sqrt{#1}}%
\global\long\def\pt{\cdot}%
\global\long\def\v#1{\mathbf{#1}}%
\global\long\def\l{\mathcal{L}}%

\def\actaa{\ref@jnl{Acta Astron.}}      % Acta Astronomica

\usepackage{xcolor}

\makeatother

\title[Deus Ex Statistica]{Deus Ex Statistica: A Statistical Solution to the binary-binary Outcome of the Chaotic, Non-Hierarchical Four-Body Problem}
\author[Y. Zack et al.]{
Yoav Zack,$^{1,2}$\thanks{E-mail: yoav.zack@weizmann.ac.il}
Hagai B.~Perets,$^{1}$\thanks{E-mail: hperets@technion.ac.il}
and Yonadav Barry Ginat$^{3,4}$\thanks{E-mail: yb.ginat@physics.ox.ac.uk}
\\
$^{1}$Department of Earth and Planetary Sciences, Weizmann Institute for Science, 234 Herzl Street, Rehovot 7610001, Israel\\
$^{2}$Faculty of Physics, Technion -- Israel Institute of Technology, Haifa 3200003, Israel\\
$^{3}$Rudolf Peierls Centre for Theoretical Physics, University of Oxford, Parks Road, Oxford, OX1 3PU, United Kingdom\\
$^{4}$New College, Holywell Street, Oxford, OX1 3BN, United Kingdom}

\date{Accepted XXX. Received YYY; in original form ZZZ}

\pubyear{2026}

\begin{document}
\label{firstpage}
\pagerange{\pageref{firstpage}--\pageref{lastpage}}
\maketitle

\begin{abstract}
We present an analytical, statistical solution to the binary-binary (2+2) outcome of the chaotic non-hierarchical four-body problem. The solution is based on the density-of-states formulation pioneered by J.~J.~Monaghan. The method skips the computationally expensive integration of the equations of motion, and instead samples the outcome from the chaotic phase-space, subject to conservation of energy, momentum, and angular momentum. From the joint distribution, we extract marginal distributions of several key parameters using Monte-Carlo integration, and numerically verify them by comparing to an identical ensemble of scattering experiments produced by the \FEWBODY ~code. From the comparison, we identify a regime not represented by the density-of-states formulation: the hard binary regime, where one binary is much harder then the four-body energy scale, and the system acts as an effective three-body system. We hypothesize that at this regime the system's probability distribution spreads over an effective reduced three-body chaotic phase space. The process of transfer from four-body to three-body phase-space is still not understood, and represents the next natural extension of density-of-states methods.
\end{abstract}

\begin{keywords}
	celestial mechanics -- chaos -- scattering -- stars: kinematics and dynamics -- galaxies: star clusters: general -- binaries: close
\end{keywords}

%%%%%%%%%%%%%%%%% BODY OF PAPER %%%%%%%%%%%%%%%%%%
\section{Introduction}\label{sec:introduction}

The Four-Body Problem (4BP) is one of the oldest problems in physics. This simple gravitational structure is ubiquitous in the universe, most notably in stellar systems and clusters as a result of binary-binary interactions \citep{sigurdssonBinarySingleStarInteractions1993, valtonenThreebodyProblem2006, merritDynamicsEvolutionGalactic2013, saslawGravitationalPhysicsStellar1985}. But even so, its dynamics are rich and unpredictable, due to its chaotic nature \citep{marchalFundamentalInstabilityGeneral1980, royPredictabilityStabilityChaos1991, arnoldMathematicalAspectsClassical2006}. Despite a long history of research, many questions about the 4BP remain open. 

In an astrophysical context, the four-body problem arises mostly in dense environments, like globular clusters \citep[e.g.,][]{heggieGravitationalMillionBody2003}, or in the evolution of quadruple stellar systems, where some massive stars reside \citep{Tokovini2014a,Tokovini2014b,Hamers_etal2015,vigna-gomezMassiveStellarTriples2021}. Here, we investigate the non-hierarchical limit---when there is no particular hierarchy of energies or angular-momenta between the 4 bodies; this limit is more relevant to binary-binary scattering interactions in clusters. The evolution of binaries in clusters is governed either by three-body interactions (binary-single encounters; \citealt{heggieBinaryEvolutionStellar1975, hutBinarysingleStarScattering1983, monaghanStatisticalTheoryDisruption1976, monaghanStatisticalTheoryDisruption1976a, anosovaDynamicalEvolutionEqualMass1986, mikkolaNumericalExplorationPhaseSpace1994, valtonenThreebodyProblem2006, samsingFormationTidalCaptures2017, samsingEccentricBlackHole2018, stoneStatisticalSolutionChaotic2019, ginatAnalyticalStatisticalApproximate2021, kolFluxbasedStatisticalPrediction2021, traniIslesRegularitySea2024,  randoforastierBinarysingleInteractionsDifferent2025}), or by binary-binary interactions \citep{Mikkola1983a,Mikkola1983b,Mikkola1984,Heggie2000,fregeauStellarCollisionsBinarybinary2004,Antognini_Thompson2016,Zevin_etal2019,marinpinaDemographicsThreebodyBinary2024,marinpinaInteractionsBinaryBlack2025}. We note, following \cite{heggieGravitationalMillionBody2003}, that binary-single scattering is essentially a three-body interaction, while binary-binary scatterings are actually a two-body interaction, albeit of two composite bodies. Thus, the latter can display entirely different dynamical behaviour. Understanding the fates of binaries in clusters is extremely important for modelling the evolution of these clusters, but also for understanding the formation of that many transients that occur there, including black-hole binaries that may form the progenitors of gravitational-wave sources seen by the LIGO-Virgo-KAGRA collaboration \citep[e.g.][]{LIGOVirgo2016,LVK_GWTC3_2023,Antoninietal2023,LVK2025GWTC4,LVK_GWTC5_2026,Weatherfordetal2020,Arcaseddaetal2023a,Ishchenkoetal2024,2019PhRvD.100d3027R,Rodriguezetal2022,Maietal2026,Ye_etal_2026,Antoninietal2025a,Ginat_etal2026}, and the formation of blue stragglers \citep{leighConstraintsBlueStraggler2019,leighAnalyticModelBlue2011}. 

Existing approaches to this problem include numerical $N$-body simulations, either of entire clusters \citep{breenDynamicalEvolutionBlack2013, barberFormationEvolutionBinary2025,ishchenkoDynamicalEvolutionMilky2024, leighWhenDoesStar2016, hurleyCompleteNbodyModel2005, gellerDIRECTNBODYMODELING2012,Arcaseddaetal2023a}, or of binary-binary scatterings in isolation \citep{Mikkola1983a,Mikkola1983b,Mikkola1984,Heggie2000,fregeauStellarCollisionsBinarybinary2004,Antognini_Thompson2016,Zevin_etal2019,marinpinaDemographicsThreebodyBinary2024,marinpinaInteractionsBinaryBlack2025,barreraretamalChaoticFourbodyProblem2024}. The former type of simulation is limited in scope, primarily due to computational complexity, whereas the latter only yields predictions for the distribution of outcomes across an ensemble of similar (in some sense) problems. One conclusion arising from these works is that for binary fractions larger than 10\%, binary-binary interactions dominate over binary-single ones \citep{sanaBinaryInteractionDominates2012, sigurdssonBinarySingleStarInteractions1993, leighAnalyticTechniqueConstraining2011}, highlighting the importance of understanding 4BPs \citep[cf.][]{Zevin_etal2019,marinpinaDemographicsThreebodyBinary2024}.

Given that simulations can only predict the distribution of outcomes---the orbital parameters of the remnant bound system (a pair of binaries, a single binary, or a hierarchical triple)---one can endeavour to formulate a theoretical prediction for such a distribution. This approach was pioneered by \cite{heggieBinaryEvolutionStellar1975, monaghanStatisticalTheoryDisruption1976, monaghanStatisticalTheoryDisruption1976a}. In particular, \cite{monaghanStatisticalTheoryDisruption1976} suggested a method for investigating the distributions directly, termed the `density-of-states formalism': the method replaces the initial condition (before chaos ensues) with an infinitesimal ensemble in phase space. This ensemble evolves dynamically in the chaotic regime of the problem, and due to the chaos, spreads evenly over all available phase-space, limited to a certain region where it mixes rapidly---the `ergodic' region. Therefore, calculating a statistically accurate outcome for the problem reduces to finding the appropriate phase-space measure on this region, subject to all relevant conservation laws. A dynamical problem is reduced to a statistical-mechanical one.

For the three-body problem, the density-of-states formalism proved extremely successful at predicting the distributions of outcome parameters and the branching ratios of different configurations \citep[e.g.][]{monaghanStatisticalTheoryDisruption1976, monaghanStatisticalTheoryDisruption1976a, heggieBinarySingleStarScatteringVII1996, valtonenThreebodyProblem2006, stoneStatisticalSolutionChaotic2019, ginatAnalyticalStatisticalApproximate2021, ginatThreebodyBinaryFormation2024,Meylakh_etal2026}. The philosophy of the method also opened a doorway to other formalisms, based on the phase-space flux out of the non-hierarchical, ergodic region (as opposed to its phase-space volume; \citealt{kolFluxbasedStatisticalPrediction2021}). It was also expanded in a natural way to include more astrophysical considerations, such as tidal forces and dynamical friction \citep{ginatAnalyticalStatisticalApproximate2021, ginatAnalyticModellingBinarysingle2022}. The 4BP, on the other hand, is still under active study \citep{nashStatisticalTheoryDisruption1980, walkerStabilityCriteriaManybody1983a, leighChaoticFourbodyProblem2016}. We highlight the work by \cite{barreraretamalChaoticFourbodyProblem2024}, which used loss-cone techniques within the density-of-states formalism to get accurate power-law estimates of various parameters for four-body interaction within the density-of-states formalism, in a similar manner to earlier works on the three-body-problem \citep{valtonenThreebodyProblem2006}. That work, however, relied on various approximations to curve the sub-space delineated by the conservation laws, mostly for the angular momentum conservation.

In this paper we extend the above works by writing the density-of-states formalism to the general, non-hierarchical 4BP. For a binary-binary scattering with a negative total energy, the allowed outcomes are \citep[e.g.,][]{arnoldMathematicalAspectsClassical2006}: a pair of binaries (2+2), a binary and two single bodies (2+1+1), and a bound, hierarchical triple plus a single unbound body (3+1); a non-hierarchical triple with an unbound single will further disintegrate into a binary and two unbound singles. We solve for the binary-binary, outcome, and obtain a high-dimensional distributions. This distribution can be converted into multiple 1D marginal distributions of the various parameters of the problems. 

For the binary-binary case, we test the theoretical predictions numerically, by comparing them with distributions obtained for analogous initial conditions from a suite of numerical simulations using the \FEWBODY ~code \citep{fregeauStellarCollisionsBinarybinary2004}. Special attention is paid to the case of an outcome consisting of a pair of binaries.

We begin in \S\ref{sec:prem}, and describe some preliminaries required for the density-of-states formalism. Next, in \S\ref{sec:22_dist}, we present the formal distribution of the 2+2 outcome, and parametrise the ergodic phase-space using a set of angular limits. The rest of the derivation appears in appendices \ref{app:22-derivation}. In \S\ref{sec:verification}, we empirically find the parameters of the phase-space, and verify the resulting outcome distribution of this case. We disucess our results in \S\ref{sec:discussion} and summarize them in \S\ref{sec:summary}.

\section{Preliminaries}\label{sec:prem}

\subsection{Hierarchy Tree}\label{subsec:hierarchy-tree}

Throughout this paper, following the density-of-states formalism, we assume that any gravitational system is either completely hierarchical or completely chaotic. If it is hierarchical (which is true well before and well after chaotic dynamics end), it can be accurately represented by a hierarchical tree.

To build such a tree, the pair of objects with the lowest energy is repeatedly inserted into the tree and replaced with their centre of mass. Denoting the new centre of mass body $j$, its mass is $M_j = m_1 + m_2$ while its reduced mass is defined using $\mathcal{M}_j = m_1 m_2 / M_j$. Note that masses of real bodies are denoted with lowercase $m$, masses of virtual bodies (i.e., centres-of-mass) are denoted with capital $M$, and reduced masses are be represented by $\mathcal{M}$.

\begin{figure}
    \centering
    \includegraphics[width=0.8\columnwidth]{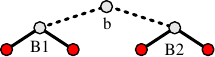}\\
    \vspace{10pt}
    \caption{The tree representing the binary-binary outcome configuration of the 4BP. Each red leaf represents a physical body, while each gray node represents the centre of mass of all of the bodies below it.}
    \label{fig:trees}
\end{figure}

In this work, we investigate the only the binary-binary outcome (the hierarchical tree representing it appears in Fig. \ref{fig:trees}). We assume that each body has a different mass while deriving theoretical distributions, and later focus on the equal mass case for the deeper investigation of the 2+2 case. 

\subsection{Delaunay Coordinates}\label{subsec:delauney_coordinates}

Modern formulations of the density-of-states formalism rely on Delaunay coordinates. This set of coordinates treats each level of the hierarchy as an independent two-body system, simplifying the representation of the phase-space immensely.

The Delaunay coordinates for a bound (elliptic) system are defined as three pairs, fitting each momenta with an upper-case letter and its corresponding angle in lower case \citep{valtonenThreebodyProblem2006}:
\begin{align}
\Lambda & =\s{GMa} & \lambda & =\lambda\\
\Gamma & =\s{GMa\left(1-e^{2}\right)} & \gamma & =\omega\\
H & =\s{GMa\left(1-e^{2}\right)}\cos I & \eta & =\Omega
\end{align}
Where $M=m_1 + m_2$ is the total mass of the effective two-body system and $\left\{ a, e, I,\Omega, \omega, \lambda\right\}$ are the standard orbital elements of the effective two-body orbit.

The Delaunay variables also exist in a hyperbolic variant \citep{floriaSimpleDerivationHyperbolic1995}, which describes two unbound bodies, i.e., in a hyperbolic trajectory:
\begin{align}
\l & = -\s{GM\left(-a\right)} & \ell & = nt\\
\mathcal{G} & = \s{GM\left(-a\right)\left(e^2-1\right)} & g & = \omega\\
\mathcal{H} & = \s{GM\left(-a\right)\left(e^2-1\right)}\cos I & h & = \Omega 
\end{align}
Where $n$ is the mean motion, i.e., the rate of change of the mean anomaly of the hyperbolic trajectory. Recall that the SMA is negative for hyperbolic orbits.

The usefulness of these coordinates is two-fold: first, they are more geometrically meaningful in cases of hierarchical structures (e.g., integration limits are a lot simpler). Second, they are a set of (almost) canonical coordinates, i.e., the Jacobian of the transformation from $\left\{ \v r,\v p\right\}$ to Delaunay variables is $\propto M^6$, which is a constant.

\subsection{The density-of-states formalism}\label{subsec:monaghan}

In order to calculate the outcome distributions for the four-body cases, we start from the general expression for the total cross section, i.e., the phase-space volume of the chaotic interaction:
\begin{align}
\label{eq:cross-section}
\sigma & =\int_{\mathcal C}\prod_{i}{\rm d}\mathbf{r}_{i}{\rm d}\mathbf{p}_{i}\d\left(\sum_{i}m_{i}\mathbf{r}_{i}\right)\d\left(\sum_{i}\mathbf{p}_{i}\right)\nonumber\\
 & \times\d\left(\sum_{j}E_{j}-E\right)\d\left(\sum_{j}\mathbf{L}_{j}-\mathbf{L}\right)
\end{align}
Where $m_i$, $\mathbf{r}_{i}$, $\mathbf{p}_{i}$ are the mass, location, and momentum of each body, $E$ and $\mathbf L$ are the total energy and angular momentum of the system, and $E_j$ and $\mathbf L_j$ represent the energy and angular momentum between a pairs of bodies, assuming the system can be described as a hierarchical structure -- in other words, the $j$ index loops over all non-leaf nodes in the hierarchical tree representing the system. Finally, the integration region $\mathcal C$ is defined as all points in phase-space which correspond to four-body chaotic interaction.

The process of converting Eq. \ref{eq:cross-section} into a valid distribution has been performed by many different methods over the years. Here, we present a simplification of the method described by \cite{stoneStatisticalSolutionChaotic2019}, composed of the following three steps:

\begin{enumerate}
    \item Change variables into Delaunay coordinates, taking care to convert hyperbolic orbits into hyperbolic coordinates, and elliptic orbits into elliptic coordinates. In this step, we implicitly assume that the system at the break-up moment can be approximated by a hierarchical structure of effective two-body systems. 

    \item Convert to the coordinates that appear in the $\d$-function (usually $E$, $L$ and $C$). This is done in order to integrate over them, and results in a set of Jacobians added to the integrals. These Jacobians are extremely important and represent most of the variability of the final distribution.
    
    \item Integrate over the delta-functions, and any over variables which are not the outcome variables required for the distribution. Integrations performed here are equivalent to tracing over dimensions of the final distribution.
\end{enumerate}

After these three steps, the leftover integrand under the integral sign simply is the distribution we are after.

\section{The binary-binary Distribution}\label{sec:22_dist}

\subsection{The Formal Distribution}\label{subsec:formal-dist}

We follow the density-of-states formalism process from \S\ref{subsec:monaghan} to derive the distribution of the binary-binary outcome. We start from the following expression for the total interaction cross section:
\begin{align}
\sigma & =M^{3}_{b}M^{3}_{B1}M^{3}_{B2}\int\left(\rmd\l_{b}\rmd\mathcal{G}_{b}\rmd\mathcal{H}_{b}\rmd\ell_{b}\rmd g_{b}\rmd h_{b}\right)\nonumber \\
 & \times\int\left(\rmd\Lambda_{B1}\rmd\Gamma_{B1}\rmd H_{B1}\rmd\lambda_{B1}\rmd\gamma_{B1}\rmd\eta_{B1}\right)\nonumber \\
 & \times\int\left(\rmd\Lambda_{B2}\rmd\Gamma_{B2}\rmd H_{B2}\rmd\lambda_{B2}\rmd\gamma_{B2}\rmd\eta_{B2}\right)\nonumber\\
 & \times\d\left(E_{{\rm B1}}+E_{{\rm B2}}+E_{b}-E\right)\nonumber \\
 & \times\d\left(L_{{\rm B1},z}+L_{{\rm B2},z}+L_{b,z}-L\right)\nonumber \\
 & \times\d\left(L_{{\rm B1},x}+L_{{\rm B2},x}+L_{b,x}\right)\nonumber \\
 & \times\d\left(L_{{\rm B1},y}+L_{{\rm B2},y}+L_{b,y}\right)
\end{align}
Where $B_1$ and $B_2$ are the parameters of the 1st and 2nd binaries, respectively, and $b$ are quantities of the hyperbolic trajectory of the binaries with respect to each other. We perform the three steps outlined in section \ref{subsec:monaghan} and get the following distribution:
\begin{align}
{\displaystyle \frac{{\rm d}\sigma}{{\rm d}B_{1}{\rm d}B_{2}}} & {\displaystyle =\sqrt 8\pi^{3}G^{3}\frac{M_{b}^{4}M_{B1}^{4}M_{B2}^{4}}{\mathcal{M}_{b}^{3/2}\mathcal{M}_{B1}^{3/2}\mathcal{M}_{B2}^{3/2}}\frac{L_{B1}L_{B2}}{\left(E_{b}E_{B1}E_{B2}\right)^{3/2}}}\nonumber\\
 & {\displaystyle \times\int\frac{\D\lambda_{B1}\D\lambda_{B2}\D\ell_{b}}{L_{b}}\rmd\eta_{B1}\rmd\eta_{B2}}
\end{align}
Where $B1$ and $B2$ represent the energy, angular momentum, and cosine inclination of each binary (B1 and B2), and $\D\ell_{b}$, $\D\lambda_{B1}$, and $\D\lambda_{B2}$ are all physical limits on the size of the chaotic region in phase-space. The distribution may appear to be non-normalizable because of an IR divergence in the energies. This is solved by the chaotic region limits $\D\lambda_i$ and $\D\ell_i$: these are cutoffs on the size and shape of the integration region, resulting from the phase-space limits on chaotic mixing. Section \ref{subsec:chaotic-region} explains their meaning. Also, there is a leftover integral over two angles. Numerically, it can be simplified to a 1D integral for faster computation, but it cannot be completely eliminated with out further assumptions.

The full derivation appears in appendix \ref{app:22-derivation}. Now we present an in-depth explanation of the meaning behind the chaotic region.

\subsection{The Chaotic Region}\label{subsec:chaotic-region}

The limits $\D\lambda_i$ and $\D\ell_i$ in the distributions parametrize the shape of the chaotic region in phase-space. This definition is an extension of the chaotic region definition by \cite{stoneStatisticalSolutionChaotic2019}, where the authors assume that the 3BP chaotic phase-space region is limited to a sphere of radius $R$ around the binary (in real space).

The limits we suggest for the 4BP work in quite the same way, and are split into \textit{hyperbolic} limits and \textit{elliptic} limits.

\subsubsection{Hyperbolic Limits}\label{subsubsec:hyperbolic-limits}

Consider two systems of bodies on a relative hyperbolic trajectory. Following \cite{stoneStatisticalSolutionChaotic2019}, we assume that the systems will interact chaotically as long as their centre-of-mass distance is smaller than a given radius $R_{\max}$, so we limit the integration region to it. Since we are working in Delaunay coordinates and distance is not explicit, we use the mean anomaly $\ell$ of the relative trajectory to parametrize it:
\begin{equation}
    \ell\left(R\right)=e\sinh\left(\cosh^{-1}\f{a-R}{ea}\right)-\cosh^{-1}\f{a-R}{ea}
\end{equation}
Since $\ell$ spans $\left[-\infty,\infty\right]$, the chaotic regions spans from $-\ell\left(R_{\max}\right)$ to $\ell\left(R_{\max}\right)$. To be consistent with later notation, we denote:
\begin{equation}
    \D\ell_{i} = \ell\left(R^{\left(i\right)}_{\max}\right)
\end{equation}
Where the $i$ index represents the $i$-th relation. The length in $\ell$ of the chaotic region is therefore $2\D\ell_{i}$. This is true for any unbound pair of systems, assuming a proper $R$ expression is found.

\subsubsection{Elliptic Limits}\label{subsubsec:elliptic-limits}

We derive the elliptic limits $\D\lambda_i$ in a similar manner to the hyperbolic ones. Consider two systems of bodies orbiting one another in an elliptic orbit. We again express a limit on the distance $\rho$ between the two systems using their relative mean anomaly $\lambda$:
\begin{equation}
    \lambda\left(\rho\right)=\cos^{-1}\f{a-\rho}{ea}-e\sin\left( \cos^{-1}\f{a-\rho}{ea}\right)
\end{equation}

As before, the chaotic region is limited by an upper limit on this distance, now called $\rho_{\max}$. But now we also add a lower limit, denoted $\rho_{\min}$. We find that this limit is necessary to remove an IR energy divergence in the distributions, by limiting from below the absolute value of the binding energy of binaries. The resulting chaotic region in mean-anomaly space is therefore:
\begin{equation}
    \D\lambda_{i}=\lambda\left(\rho^{\left(i\right)}_{\max}\right) - \lambda\left(\rho^{\left(i\right)}_{\min}\right)
\end{equation}
But since the mean anomaly is symmetric to negation, the actual integration region is doubled, and is of measure $2\D\lambda_i$.

To understand that lower limit, consider the case of two interacting binaries: for complete four-body chaos, we must require that the binary sizes are not too different from their relative centre-of-mass distance, and from each other. Otherwise we may stray into an effective three-body regime, where one binary is hard enough to be treated as an effective single with the combined mass of the other two.

\subsection{Explicit limits}\label{subsec:22-limits}

In order to define $R$, i.e., the maximal distance between two binaries for which they still interact chaotically, we start from the simpler case of the 3BP. Assume a binary denoted ``B", and denote ``s" its relationship with another "outer" body is denoted by ``s" (e.g., the relative energy between the binary and the single is $E_s$). The limit $R$ in this case can be taken directly from \cite{ginatAnalyticalStatisticalApproximate2021}. We do, however, replace $a$ with the binary's radius averaged over the mean anomaly, $\left\langle r\right\rangle_{M}$, as numerical testing proved it a better fit. This average is given by
\begin{equation}
    \left\langle r\right\rangle_{M} = a\left(1+\frac{e^2}{2}\right),
\end{equation}
and represents the time-averaged distance of the escaping orbit. We use this, rather than $a$, because this accounts for the longer time spent further away, while still acknowledging that the system might not disintegrate at apoapsis. 

We get the following definition for the 3BP chaotic radius:
\begin{equation}
\label{eq:original-R}
R_{\max}=\beta\min\left\{ \sqrt[3]{\f{G\mathcal{M}_{B}\mathcal{M}_{s}M_{s}}{M_{B}\left|E\right|}\left\langle r_{B}\right\rangle _{M}^{2}},\left\langle r_{B}\right\rangle_{M}\right\}
\end{equation}
Next, we extend this definition to account for the fact that the ejected body is a binary $Bj$. We treat it as a sphere of apoapsis radius, and demand that all the sphere will be outside distance $R$ from the original binary $Bi$. This adds the following component to $R$:
\begin{align}
R_{\max} & =\beta\min\left\{ \sqrt[3]{\f{G\mathcal{M}_{Bi}M_{b}}{M_{Bi}\left|E\right|}\left\langle r_{Bi}\right\rangle ^{2}_{M}},\left\langle r_{Bi}\right\rangle _{M}\right\} \nonumber \\
 & +\beta\f{\max\left\{ m_{1Bj},m_{2Bj}\right\} }{M_{Bj}}\left\langle r_{Bj}\right\rangle _{M}
\end{align}
Finally, since we now have a symmetry between the bodies (each can be the ejected one), we define two symmetric chaotic radii as follows:
\begin{align}
R^{\left(Bi\right)}_{\max} & =\beta\min\left\{ \sqrt[3]{\f{G\mathcal{M}_{Bi}M_{b}}{M_{Bi}\left|E\right|}\left\langle r_{Bi}\right\rangle ^{2}_{M}},\left\langle r_{Bi}\right\rangle _{M}\right\} \nonumber \\
 & +\beta\f{\max\left\{ m_{1Bj},m_{2Bj}\right\} }{M_{Bj}}\left\langle r_{Bj}\right\rangle _{M}
\end{align}
In order to get a single $R_{\max}$, one may consider using $R_{\max}=\max\left\{ R^{\left(B1\right)}_{\max}, R^{\left(B2\right)}_{\max}\right\}$, as this is the radius at which at the last chaotic interaction before escape -- i.e., the last body of binary $i$ gets outside the chaotic radius of binary $j$.

But this line of thinking is erroneous: the density-of-states formalism assumes that the final interaction before escape leaves the probability distribution evenly spread over all of the chaotic 4BP phase-space, not a 3BP phase-space. So, we must demand that at that final interaction, \textit{all} bodies were in chaotic interaction, not just the last three. Therefore, we define:
\begin{equation}
    R_{\max}=\min\left\{ R^{\left(B1\right)}_{\max},R^{\left(B2\right)}_{\max}\right\} \label{eq:hyperbolic_R}
\end{equation}
This way, we demand that at the last scattering before escape, the interaction is fully 4-body chaotic.

All of these considerations completely define the hyperbolic limit $\D\ell$. Now, we turn to the elliptic limits $\D\lambda_{i}$. Finding appropriate expressions for $\rho_{\min}$ and $\rho_{\max}$ for hyperbolic edges is a difficult problem, as it requires a deep understanding of the dynamics of the problem.

We tested several options for $\rho$ by comparing the resulting marginal distributions of the 2+2 parameters to the same distributions given by 2+2 numerical simulations (see Sec \ref{subsec:FEWBODY}). The tests included limits based on binary SMA, eccentricity, energy, or the hyperbolic $R$. Most didn't result in a good fit, but some did, and the best-fitting limits for the 2+2 elliptic limits are provided here:
\begin{align}
    \rho_{\max, Bi} & = -\frac{1}{4}\frac{G M_{Bi}^2}{E} \\
    \rho_{\min, Bi} & = \alpha R_{\max}
\end{align}
Where $M$ is the total mass of the interacting bodies (i.e., the total mass of the system), and $\alpha\simeq0.01$ is a numerical factor calibrated empirically. These two definitions are by no means final, and we hope future works will improve upon them.

\section{Numerical Verification of the binary-binary Case}\label{sec:verification}

Here we present a comparison of the 2+2 distribution to numerical simulations. The comparison is made via the various marginal distributions resulting from the density-of-state formalism. The integration of the complete disturbance is performed via MC integration, and the resulting marginal is than compared with the same marginal extracted from an ensemble of \FEWBODY~numerical simulations.

\subsection{MC Integration}\label{subsec:mc-integration}

\begin{figure}
     \centering
     \includegraphics[width=0.7\columnwidth]{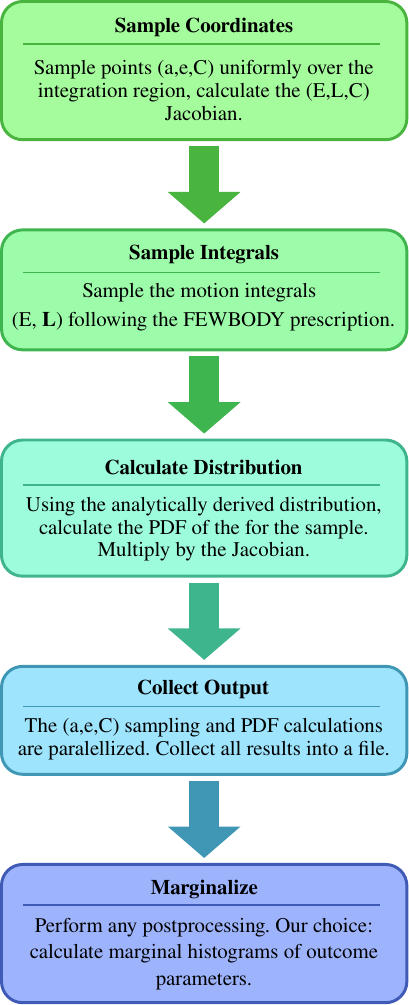}
     \caption{The processing pipeline for the numerical integration of the distribution, using MC integration.}
     \label{fig:mc_pipeline}
\end{figure}

We begin with the integration process, where a schematic analysis of the MC integration pipeline appears in Fig. \ref{fig:mc_pipeline}.

The formal distribution given in \S\ref{sec:22_dist} is given in $\left(E, L, C\right)$ coordinates, which are convenient to work with analytically. But numerically, two of the three coordinates are unbounded, and sampling them randomly (as required for MC integration) will not necessarily result in a binary-binary system. This can be fixed by rejecting incorrect samples, but this lowers integration efficiency. We choose instead to sample in $\left(e, a, C\right)$ coordinates, and multiply the distribution by the following Jacobian:
\begin{align}
J_{Bi}^{\rm (eaC)} = \frac{\mathcal M_{Bi}^{2}}{2}\left(\frac{G M_{Bi}}{a_{Bi}}\right)^{\frac32} \frac{e_{B_i}}{\sqrt{1-e_{Bi}^2}}
\label{eq:aec_jacobian}
\end{align}
This allows two of the three coordinates to be bound, and forces all sampled points to represent a binary-binary configuration.

Using this method, we sample points for which we will calculate the distribution. In total, we sampled $2^{24}$ points per CPU core, and used $4$ nodes, each with $96$ cores, to a total of about $6.4\times 10^{9}$ samples. Since the distributions include an inner integration, each sample gets a grid of $31\times31$ points in the range $\left[0, 2\pi\right]^2$ representing $\eta_{Bi}$ values, and the inner integrand of the distribution is calculated over them.

The distribution value is then calculated for all points, multiplied by the Jacobians, and returned. Note that some points evaluated to probability zero, of the sampled configuration was outside of the chaotic phase--space, as defined in \ref{subsec:chaotic-region}.

\subsection{\FEWBODY\ Simulations}\label{subsec:FEWBODY}

\begin{figure*}
    \centering
    \includegraphics[width=\textwidth]{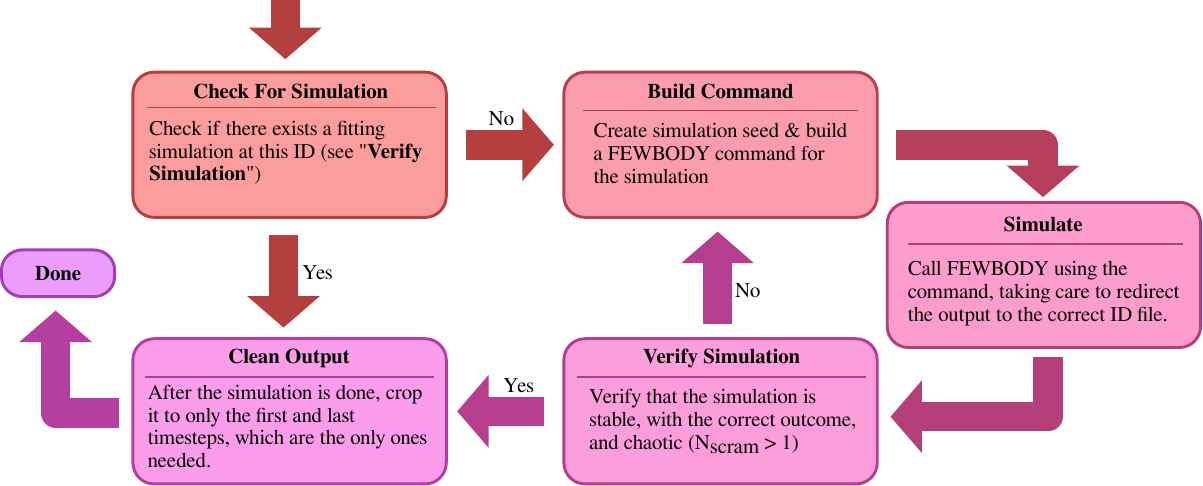}
    \caption{The simulation pipeline of the \FEWBODY simulations. Note that this loop runs for each simulation ID until all are done.}
    \label{fig:fewbody_pipeline}
\end{figure*}

The baseline truth we compare our theoretical predictions to is a set of numerical simulations performed using the \FEWBODY ~code \citep{fregeauStellarCollisionsBinarybinary2004}. We selected \FEWBODY ~for several reasons, among them its reliability, ease of use, and its established status for small N-body simulations for the past 20 years.

For each of the cases we investigate, we generated a random ensemble of 12,288 (384 cores, 32 simulations per core) simulations with known seeds, following the prescription shown in Fig. \ref{fig:fewbody_pipeline}. These simulations were used as a database. 

The simulation process (similarly to the MC integration) uses a variation of rejection sampling: to get a chaotic simulation to end in the binary-binary configuration, we perform simulations with random initial conditions until the outcome fits the requirements, which are:

\begin{enumerate}
    \item The simulation is finished, i.e., the final configuration is stable (according to \FEWBODY's criteria; see \citealt{fregeauStellarCollisionsBinarybinary2004}).
    \item The simulation's outcome has the required hierarchical structure -- a binary-binary, in this case.
    \item The simulation is chaotic, i.e.$N_{\rm scram} \geq 2$ where $N_{\rm scram}$ is the scramble number defined by \FEWBODY~\citep{fregeauStellarCollisionsBinarybinary2004, stoneStatisticalSolutionChaotic2019}.
\end{enumerate}

If a simulation does not meet all three criteria, we reject it, generate a new seed, and run a new simulation. The seed is a 15-digit integer (the maximal size of seed allowed in \FEWBODY) generated randomly using \texttt{Numpy}. This process is used for each simulation until the database is complete.

From the database, we extracted histograms of various relative parameters of the bodies: their relative angular momentum, energy, eccentricity, inclination, etc. In order to generate the histograms, we created a general \texttt{Python} suite that analyses any ensemble of simulations from \FEWBODY.

The suite works as follows: first, it homogenizes the different hierarchical trees, by pinning the most deeply nested pair as the lowest level of the tree, ordering the edges of the tree by energy, and orienting each simulation based on the direction of its angular momentum vector (so it points towards the positive $z$ axis). Next, the virtual bodies (COMs) are calculated based on the tree, and the required parameter (angular momentum, energy, etc.) is calculated for each of them -- i.e., between the two child nodes composing the COM in question. The resulting quantities for each virtual body are then collected into a 1D histogram, one for each virtual body (So in total $N-1=3$ histograms).

\subsection{Results}\label{subsec:22_results}

\begin{figure*}
    \centering
    \includegraphics[width=0.45\linewidth]{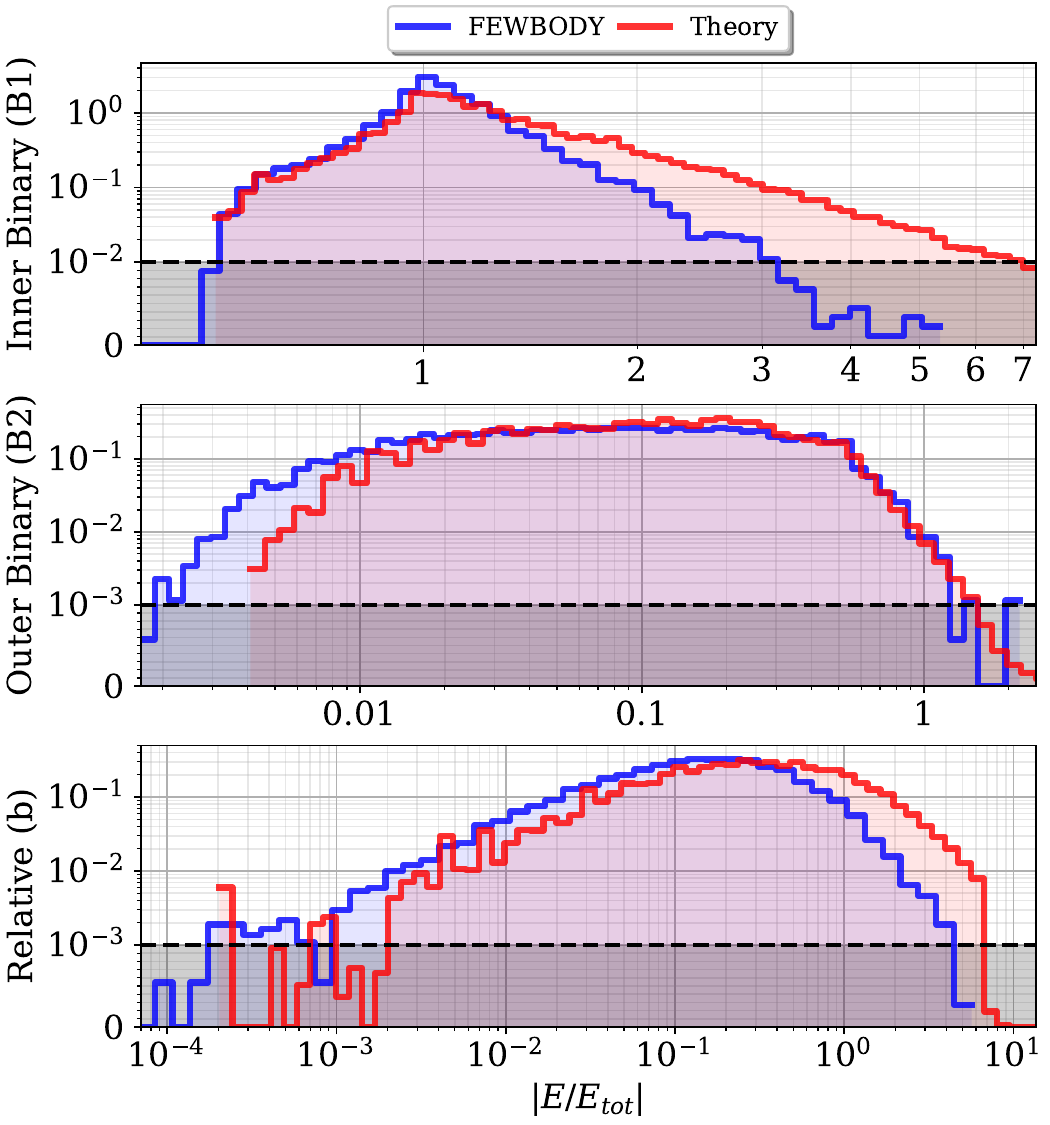}
    \includegraphics[width=0.45\linewidth]{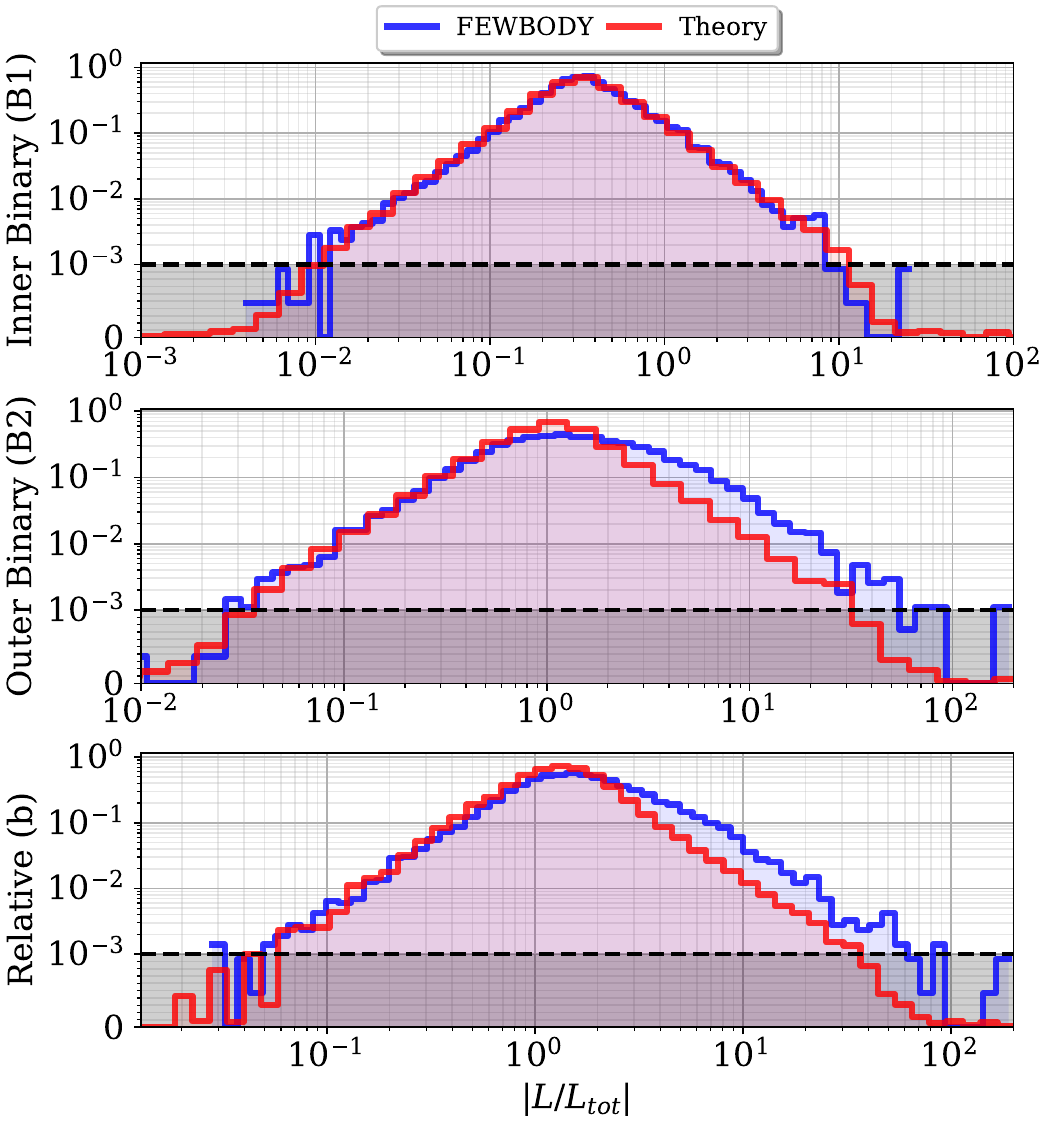}
    \includegraphics[width=0.45\linewidth]{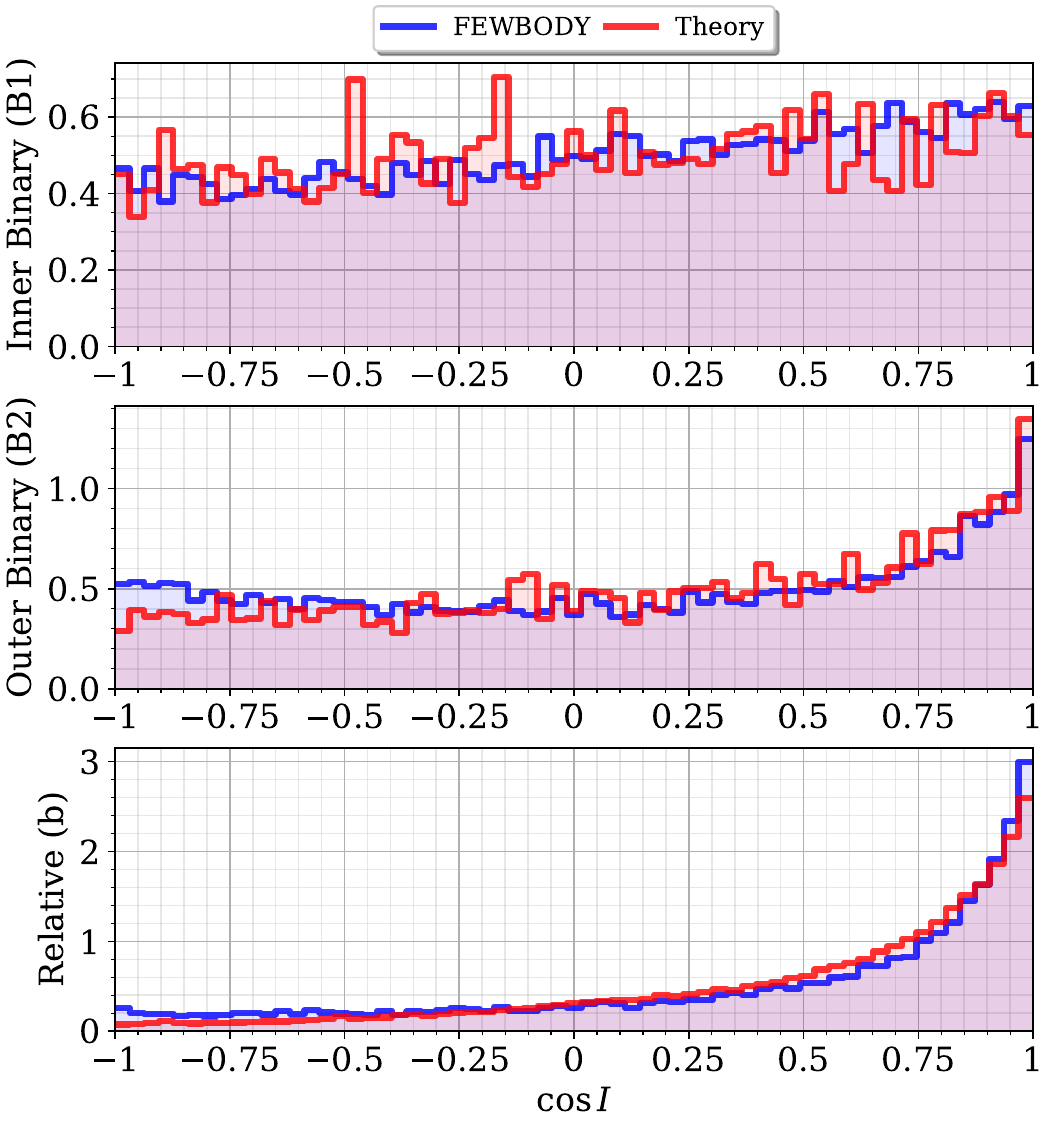}
    \includegraphics[width=0.45\linewidth]{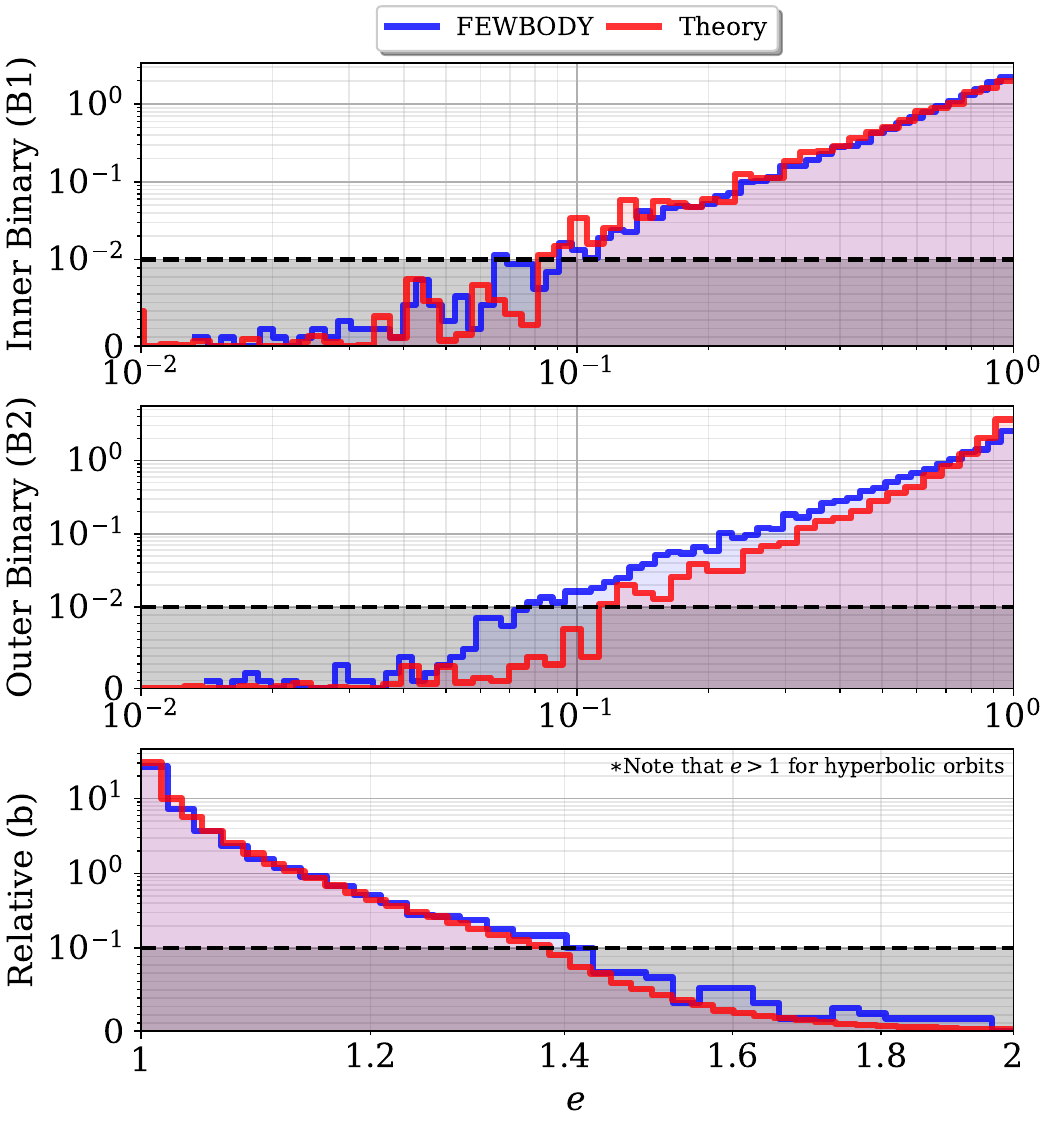}
    \caption{A selection of the results of the marginal-distribution comparison between the 2+2 theory (MC integration) and simulations. In each panel, the 1st plot represents the tight binary, the 2nd one is the wide binary, and the 3rd is their relative orbit. Top left panel: energy. Top right panel: angular momentum. Bottom left panel: cosine inclination. Bottom right panel: eccentricity. Disregarding noise, there is a clear fit between theory and simulations. Note that since the histograms are with log-scaled bins, a distribution with power law $\alpha$ will be represented as a linear line with slope $\alpha+1$. For the same reason, we note that we are plotting $P\left(\log x\right)$ in each plot. The $log$ part was removed from the label to reduce clutter and improve legibility.}
    \label{fig:22_results}
\end{figure*}

Here we present the results of our investigation into the equal-mass scenario of the 2+2 outcome. A selection of the results appears in Fig. \ref{fig:22_results}. Note the different sources of noise in the figure: the noise in the \FEWBODY line stems from the limited number of simulations, while the noise in the theory line is a natural product of MC integration.

We note that most distributions demonstrate generally good fits, except for the energy distribution at the hard binary limit. There we observe a discrepancy in the power law, the theory giving a shallower power law than the \FEWBODY ~ensembles.

We believe that the source of this discrepancy is the definition of the chaotic radii. The definitions we used are relatively simple, and the distributions are highly sensitive to the specific expressions used. We believe that the issue is located in $\rho_{\min, B1}$, specifically, as this parameter affects most strongly the hard power law in B1. we explore this hypothesis in \S\ref{sec:discussion}.

\section{Discussion}\label{sec:discussion}

The analytical distributions displayed in \S \ref{sec:22_dist} describe the numerical ones in \S \ref{sec:verification} well, suggesting that the theory is generally verified. There is one significant discrepancy, however, in the hard limit of the tighter binary, where the power law predicted by the analytical distribution is incompatible with the \FEWBODY simulations. This case corresponds to the tighter binary being ejected with significantly more binding energy than the wider one. It is reasonable to expect that prior to the disintegration of the quadruple system, these two particles of the binary were still a close-knit pair, whose centre of mass was in a democratic resonance with the other two particles---at least in some fraction of the cases. It may therefore be the case that the quadruple did not have access to the full non-hierarchical region of a four-body phase space, but rather to an effective three-body sub-space.

We therefore hypothesize that the discrepancy results from cases where the tighter binary is hard enough to be treated as a single body. Our lower limit for the binary size, i.e.~$\rho_{B1}$, may not represent the true phase space accurately in this case, and includes cases where the system is only 3-body chaotic and not full 4-body chaotic. We do note, however, that there are more possibilities: since both theory and simulations may include or miss cases which are completely chaotic, there are four possible inconsistencies:

\begin{enumerate}
    \item The theoretical prediction includes too many hard binaries, which artificially raise the probability distribution (red at Fig. \ref{fig:22_results}) at the hard binary limit.
    \item The theoretical prediction includes too few soft binaries, which artificially lower the probability distribution (red at Fig. \ref{fig:22_results}) at the soft binary limit.
    \item Our filtering of \FEWBODY~ simulations includes too many soft binaries, which artificially raise the numerical histogram (blue at Fig. \ref{fig:22_results}) at the soft binary limit.
    \item Our filtering of \FEWBODY~ simulations includes too few hard binaries, which artificially lower the numerical histogram (blue at Fig. \ref{fig:22_results}) at the hard binary limit.
\end{enumerate}

We support interpretation (i), because of the natural way four-body systems may become effectively three-bodied. 

\begin{figure}
    \centering
    \includegraphics[width=0.95\linewidth]{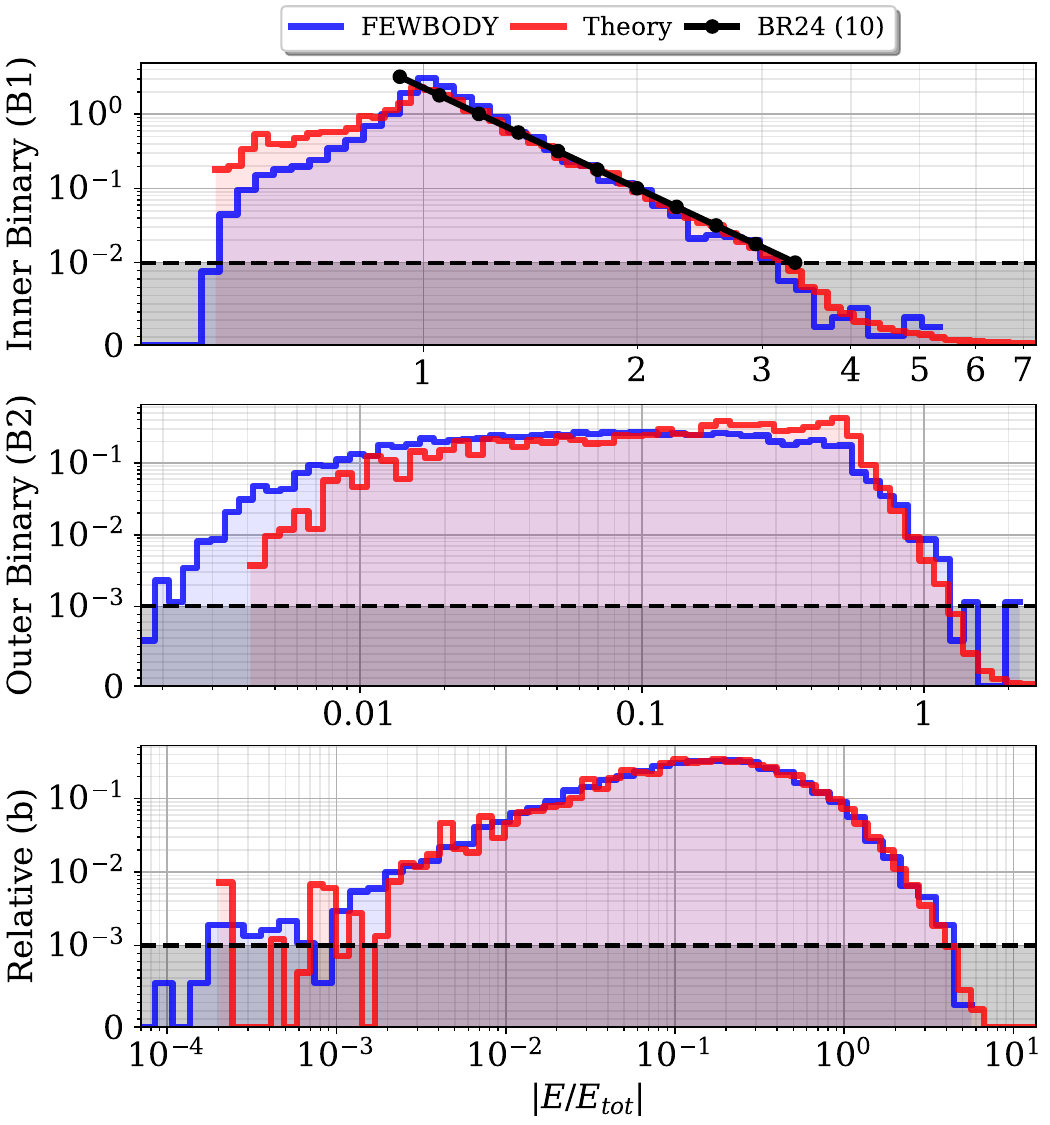}
    \caption{The 2+2 energy distribution, with the power-law fix we detected, and with an overlay of the relevant BR24 prediction. Note that again this is a log-log distribution plot (read the caption of figure \ref{fig:22_results} for the implications.)}
    \label{fig:fixed_22}
\end{figure}

Another reason to support this interpretation is the shape of the deviation: empirically, the theoretical results appear to be missing a $E_{B1}^{-2}$ factor in the range $E_{B1}>E_{\rm tot}$. In figure \ref{fig:fixed_22} we show, that if we multiply the differential probability by it, i.e.
\begin{equation}
    \frac{\mathrm{d}\sigma}{\mathrm{d}B_1 \mathrm{d}B_2} \mapsto \frac{\mathrm{d}\sigma}{\mathrm{d}B_1 \mathrm{d}B_2}\times \begin{cases}
        1 & \mbox{if } E_{B1}\leq E_{\rm tot}, \\ 
        \frac{E_{\rm tot}^2}{E_{B1}^2} & \mbox{if } E_{B1} > E_{\rm tot}
    \end{cases}\,, 
\end{equation}
this fixes the discrepancy. We also show, in the same plot, we get the same power law as the tight-binary prediction of \citet[][BR24]{barreraretamalChaoticFourbodyProblem2024} (for both binaries), which also supports our results, and this fix.

Such an expression (as a proportionality coefficient) can be achieved by appealing to ideas from loss-cone theory: for equal masses, $a_{B_1}\propto |E_{B_1}|^{-1}$. If the final scattering results in an outer scale that is not proportional to $a_{B_1}$ (i.e., the tighter binary is effectively a single), the angular size of the internal resolved region is $\theta_{\rm lc}\sim a_{B_1}/R_{\rm vir}$, where $R_{\rm vir} = G\sum_{i\neq j}m_im_j/(4\left\vert E_{\rm tot}\right\vert)$ is the typical distance between the bodies in a non-hierarchical state of energy equipartition, and the corresponding resolved fraction scales like $\theta_{\rm lc}^2$. We therefore divide the pure distribution by an area factor of $\left|E_{B_1}/E_{\rm tot}\right|^2$. We emphasise, however, that this was a heuristic, \emph{ad hoc} multiplication, and we do not contend that loss-cone theory provides the correct interpretation to the origin of this missing $E_{B_1}^2$ factor; indeed, as we remarked earlier, we believe that it stems from incorrectly characterising the ergodic sub-space that the system has access to.

Historically, a similar issue arose already in the context of the three-body problem, in the book of \citet[][pp.~174--175]{valtonenThreebodyProblem2006}. There, they presented the density-of-sates method for the three-body problem, but without an exact conservation of angular momentum. They used loss-cone arguments to try to approximate accounting for angular-momentum constraints (leading to a factor of $E_{B}^2$, just like here), and to an $E_{B}^{-9/2}$ energy distribution that agreed with Heggie's detailed-balance argument \citep{heggieBinaryEvolutionStellar1975}, a better approximations for the true distributions than without the loss-cone term. When \cite{stoneStatisticalSolutionChaotic2019} accounted for angular-momentum conservation exactly, it was seen that the loss-cone argument was redundant, and does not represent the true physical mechanism at play---the resolution lay in identifying the ergodic sub-set correctly.\footnote{Indeed, the detailed-balance arguments in \cite{heggieBinaryEvolutionStellar1975} also only conserved energy explicitly, but \cite{ginatAnalyticalStatisticalApproximate2021} showed, \emph{inter alia}, that when one does account for angular momentum in detailed balance, one does get exactly the same differential cross-section as in the density-of-states formalism of \cite{stoneStatisticalSolutionChaotic2019}.} It is our hope that, in a similar manner, future work will delineate this region better. 

To further support this speculation, recall that apart from the (2+2) case, there are more possible outcomes for the 4BP. Using the density-of-states formalism, we derived the outcome distributions of the (2+1+1) and (3+1) outcomes. The numerical verification of these distributions is much less mature compared with the (2+2) case, and even though the initial results are promising, they are not in good agreement with the simulations. The reason, we hypothesize, is that these outcome states admit large non-ergodic, albeit non-hierarchical, regions in the phase-space allowed by the conserved quantities; such states cannot be captured by density-of-states formalism straightforwardly. For example, \cite{barreraretamalChaoticFourbodyProblem2024} modelled the (2+1+1) state was modelled as a sequential escape, implying mixing only in a three-body-like phase sub-space of the full four-body phase space.

\section{Summary}\label{sec:summary}
In this work, we derived and verified a complete semi-analytical solution to the binary-binary outcome of the four-body problem. The derivation is based on the density-of-states formalism, i.e., it assumes that at the break-up time, the system is at the boundary between chaotic and hierarchical interaction, and integrates over the chaotic phase-space. The outcome distribution shape depends on six parameters per distribution: the angular lower and upper limits of the chaotic region of phase space, denoted $R$ (for each binary) and $\rho$ (for the distance between the binaries).

We derived expressions for these limits for the 2+2 case, getting a complete analytical distribution, up to a simple numerical integral. In order to verify it, we used an ensemble of simulations produced using the \FEWBODY ~code, and compared marginal distributions of parameters between the analytical derivation and the numerical ensemble.

We discussed (in  \S\ref{sec:discussion} above) the discrepancy at the hard binary limit, and our belief that including a better filter on the \FEWBODY ~results will result in better agreement, as has been true for the three-body-problem. We believe future work should focus on that avenue in order to complete the (2+2) verification.

Finally, based on the previous hypotheses, we also hypothesise that the density-of-states formalism is a viable method of solving any chaotic enough interaction, as long as it is possible to treat the interaction as a hierarchical tree of chaotic interactions. In such a case, each component may be solved using the required density-of-states distribution. The process repeats until a complete disintegration is achieved. Of course, this method is not practical for high $N$ problems, but we believe it can be used at least for the 5 and 6 body problems, utilising a recursive expression for the density-of-states formalism.

\section*{Acknowledgements}
We ran the simulations on the \texttt{Nyx} computing cluster at the Technion. This work was supported by the Science and Technology Facilities Council grant No.~ST/W000903/1 and by a Leverhulme Trust International Professorship Grant (No.~LIP-2020-014, PI: S.~L.~Sondhi). Y.B.G.'s work was supported in part by the Simons Foundation via a Simons Investigator Award to A.A.~Schekochihin (No.~930121).

\section*{Data Availability}
The data underlying this article will be shared upon reasonable request to the corresponding author. Codes produced for this paper can be found in the project's \href{https://codeberg.org/yoavzack/master}{repository}. \FEWBODY ~\citep{fregeauStellarCollisionsBinarybinary2004} may be obtained from the above link, or on this \href{https://sourceforge.net/projects/fewbody/files/}{link}.

\bibliography{references}

\appendix
\section{The Three Body Problem -- A Reanalysis}\label{app:21-derivation}

In this section, we describe a streamlined derivation of the 3BP distribution. This process is the basis of the 4BP derivation. We, as usual in the density-of-states formalism, start with an expression for the total chaotic phase-space volume of the 3BP:
\begin{align}\sigma & =M^{3}_{B}M^{3}_{s}\int\rmd\Lambda_{B}\rmd\Gamma_{B}\rmd H_{B}\rmd\lambda_{B}\rmd\gamma_{B}\rmd\eta_{B}\\
 & \times\int\rmd\mathcal{L}_{s}\rmd\mathcal{G}_{s}\rmd\mathcal{H}_{s}\rmd\ell_{s}\rmd g_{s}\rmd h_{s}\nonumber\\
 & \times\d\left(E_{B}\left(\Lambda_{B}\right)+E_{s}\left(\mathcal{L}_{B}\right)-E\right)\nonumber\\
 & \times\d\left(L_{B,z}\left(H_{B}\right)+L_{s,z}\left(\mathcal{H}_{s}\right)-L\right)\nonumber\\
 & \times\d\left(L_{B,x}\left(\Gamma_{B},H_{B},\eta_{B}\right)+L_{s,x}\left(\mathcal{G}_{s},\mathcal{H}_{s},h_{s}\right)\right)\nonumber\\
 & \times\d\left(L_{B,y}\left(\Gamma_{B},H_{B},\eta_{B}\right)+L_{s,y}\left(\mathcal{G}_{s},\mathcal{H}_{s},h_{s}\right)\right)\nonumber
\end{align}
We transform the integration variables to the $\d$-function ones, using the following transformations and Jacobians:
\begin{align}
\rmd\l_{s}\rmd\mathcal{G}_{s}\rmd\mathcal{H}_{s}\rmd h_{s} & =\frac{-GM_{s}}{\left(2\mathcal{M}_{s}E_{s}\right)^{3/2}L_{s}}\rmd E_{s}\rmd L_{s,x}\rmd L_{s,y}\rmd L_{s,z}\\
\rmd\Lambda_{B}\rmd\Gamma_{B}\rmd H_{B} & =\f{-GL_{B}M_{B}}{\left(-2\mathcal{M}_{B}E_{B}\right)^{3/2}}\rmd E_{B}\rmd L_{B}\rmd C_{B}
\end{align}
Where $\mathcal{M}_B=\frac{m_{1B}m_{2B}}{M_B}$ is the reduced mass of the binary and $\mathcal{M}_s=\frac{M_B m_s}{M_s}$ is the reduced mass between the binary and the single. The first transformation is chosen since $E_s$ and the components of $\mathbf{L}$ appear in the $\delta$-functions and will be used to integrate them. The 2nd transformation is chosen as the final coordinates of the resulting distribution.

Substituting these into the integral, we get:
\begin{align}\sigma & =M^{3}_{B}M^{3}_{s}\int\left(\rmd E_{B}\rmd L_{B}\rmd C_{B}\rmd\lambda_{B}\rmd\gamma_{B}\rmd\eta_{B}\right)\nonumber\\
 & \times\int\left(\rmd E_{s}\rmd L_{s,x}\rmd L_{s,y}\rmd L_{s,z}\rmd\ell_{s}\rmd g_{s}\right)\nonumber\\
 & \times\d\left(E_{B}+E_{s}-E\right)\nonumber\\
 & \times\d\left(L_{B,z}\left(L_{B},C_{B}\right)+L_{s,z}-L\right)\nonumber\\
 & \times\d\left(L_{B,x}\left(L_{B},C_{B},\eta_{B}\right)+L_{s,x}\right)\nonumber\\
 & \times\d\left(L_{B,y}\left(L_{B},C_{B},\eta_{B}\right)+L_{s,y}\right)\nonumber\\
 & \times\f 18G^{2}\f{M_{B}M_{s}}{\mathcal{M}^{3/2}_{B}\mathcal{M}^{3/2}_{s}}\f{L_{B}}{L_{s}}\f 1{\left(-E_{B}E_{s}\right)^{3/2}}
\end{align}
This form allows us to integrate over the $s$ variables $E_s$, $\v L_s$ utilizing the $\d$-functions:
\begin{align}
\sigma & =M^{3}_{B}M^{3}_{s}\int\left(\rmd E_{B}\rmd L_{B}\rmd C_{B}\rmd\lambda_{B}\rmd\gamma_{B}\rmd\eta_{B}\right)\left(\rmd\ell_{s}\rmd g_{s}\right)\nonumber\\
 & \times\f{1}8G^{2}\f{M_{B}M_{s}}{\mathcal{M}^{3/2}_{B}\mathcal{M}^{3/2}_{s}}\f{L_{B}}{L_{s}}\f 1{\left(-E_{B}E_{s}\right)^{3/2}}
\end{align}
Next, we integrate over the angles. $\lambda_{B}$, $\gamma_{B}$
and $g_{s}$ each give $2\pi$, while $\ell_{s}$ gives $2\ell_{\max}^{\left(s\right)}$:
\begin{align}
\sigma & =\left(2\pi\right)^{3}M^{3}_{B}M^{3}_{s}\int\left(\rmd E_{B}\rmd L_{B}\rmd C_{B}\rmd\eta_{B}\right)\nonumber\\
 & \times\f 18G^{2}\f{M_{B}M_{s}}{\mathcal{M}^{3/2}_{B}\mathcal{M}^{3/2}_{s}}\f{L_{B}}{L_{s}}\f 1{\left(-E_{B}E_{s}\right)^{3/2}}\pt2\ell^{\left(s\right)}_{\max}
\end{align}
Where $\ell_{\max}^{\left(s\right)}$ is the maximum of the hyperbolic mean anomaly. We simplify the expression to get:
\begin{align}
\f{\rmd\sigma}{\rmd E_{B}\rmd L_{B}\rmd C_{B}} & =\left(\sqrt{2}\pi G\right)^{2}\f{M^{4}_{B}M^{4}_{s}}{\mathcal{M}^{3/2}_{B}\mathcal{M}^{3/2}_{s}}\f{L_{B}}{\left(-E_{B}E_{s}\right)^{3/2}}\nonumber\\
 & \times\int\f{\pi\ell^{\left(s\right)}_{\max}}{L_{s}}\rmd\eta_{B}
 \label{eq:21-general}
\end{align}
In the 4BP case below this is where we stop, as it is impossible to integrate any further analytically there. But here, we can eliminate the final $\eta_{B}$ integration by utilizing the fact that
$\frac{\ell_{\max}^{\left(s\right)}}{L_{s}}$ depends on $\eta_{B}$ only through $L_{S,x}^{2}+L_{S,y}^{2}$, which is conveniently independent of $\eta_{B}$. Thus, we can integrate  over it as well and just get another factor of $2\pi$:
\begin{equation}
f_{2+1}\left(B\right)=4\pi^{4}G^{2}\f{M_{B}^{4}M_{s}^{4}}{\mathcal{M}_{B}^{3/2}\mathcal{M}_{s}^{3/2}}\f 1{\left(-E_{B}E_{s}\right)^{3/2}}\f{L_{B}}{L_{s}}\ell_{\max}^{\left(s\right)}
\end{equation}
And this is indeed equivalent to the analytical result from \cite{stoneStatisticalSolutionChaotic2019}.

\section{The Four Body Problem -- The Binary-Binary Case}\label{app:22-derivation}

The analysis of the 4BP outcomes begins similarly to the three-body case. Therefore, we add a streamlined version of the derivation in appendix \ref{app:21-derivation}. We start from the expression for the full cross-section, written in Delaunay coordinates:
\begin{align}
\sigma & =M^{3}_{b}M^{3}_{B1}M^{3}_{B2}\int\left(\rmd\l_{b}\rmd\mathcal{G}_{b}\rmd\mathcal{H}_{b}\rmd\ell_{b}\rmd g_{b}\rmd h_{b}\right)\nonumber\\
 & \times\int\left(\rmd\Lambda_{B1}\rmd\Gamma_{B1}\rmd H_{B1}\rmd\lambda_{B1}\rmd\gamma_{B1}\rmd\eta_{B1}\right)\nonumber\\
 & \times\int\left(\rmd\Lambda_{B2}\rmd\Gamma_{B2}\rmd H_{B2}\rmd\lambda_{B2}\rmd\gamma_{B2}\rmd\eta_{B2}\right)\nonumber\\
 & \times\d\left(E_{{\rm B1}}+E_{{\rm B2}}+E_{b}-E\right)\d\left(L_{{\rm B1},z}+L_{{\rm B2},z}+L_{b,z}-L\right)\nonumber\\
 & \times\d\left(L_{{\rm B1},x}+L_{{\rm B2},x}+L_{b,x}\right)\d\left(L_{{\rm B1},y}+L_{{\rm B2},y}+L_{b,y}\right)
\end{align}

Next, we define the following coordinate transformation. Note that the 1st transformation corresponds to the variables we want to eliminate using the $\d$-functions, while the 2nd one to the variables we want to keep (to be the variables in our distribution):
\begin{align}
\rmd\l_{b}\rmd\mathcal{G}_{b}\rmd h_{b}\rmd\mathcal{H}_{b} & =\f{GM_{b}}{\left(2\mathcal{M}_{b}E_{b}\right)^{3/2}L_{b}}\rmd E_{b}\rmd L_{b,x}\rmd L_{b,y}\rmd L_{b,z}\\
\rmd\Lambda_{Bi}\rmd\Gamma_{Bi}\rmd H_{Bi} & =\f{GM_{Bi}L_{Bi}}{\left(-2\mathcal{M}_{Bi}E_{Bi}\right)^{3/2}}\rmd E_{Bi}\rmd L_{Bi}\rmd C_{Bi}
\end{align}
This results in the following simplified integral, where in each $\D$-function the $b$ variable is an integration variable:
\begin{align}
\sigma & =M^{3}_{b}M^{3}_{B1}M^{3}_{B2}\int\left(\rmd\l_{b}\rmd\mathcal{G}_{b}\rmd\mathcal{H}_{b}\rmd\ell_{b}\rmd g_{b}\rmd h_{b}\right)\\
 & \times\int\left(\rmd\Lambda_{B1}\rmd\Gamma_{B1}\rmd H_{B1}\rmd\lambda_{B1}\rmd\gamma_{B1}\rmd\eta_{B1}\right)\nonumber\\
 & \times\int\left(\rmd\Lambda_{B2}\rmd\Gamma_{B2}\rmd H_{B2}\rmd\lambda_{B2}\rmd\gamma_{B2}\rmd\eta_{B2}\right)\nonumber\\
 & \times\d\left(E_{{\rm B1}}+E_{{\rm B2}}+E_{b}-E\right)\d\left(L_{{\rm B1},z}+L_{{\rm B2},z}+L_{b,z}-L\right)\nonumber\\
 & \times\d\left(L_{{\rm B1},x}+L_{{\rm B2},x}+L_{b,x}\right)\d\left(L_{{\rm B1},y}+L_{{\rm B2},y}+L_{b,y}\right)\nonumber\\
 & \times\f{GM_{b}}{\left(2\mathcal{M}_{b}E_{b}\right)^{3/2}L_{b}}\f{GM_{B1}L_{B1}}{\left(-2\mathcal{M}_{B1}E_{B1}\right)^{3/2}}\f{GM_{B2}L_{B2}}{\left(-2\mathcal{M}_{B2}E_{B2}\right)^{3/2}}\nonumber
\end{align}
We now integrate over $\mathbf L_{b}$ and $E_{b}$ and eliminate all $\d$-functions:
\begin{align}
\sigma & =M^{3}_{b}M^{3}_{B1}M^{3}_{B2}\int\left(\rmd E_{B1}\rmd L_{B1}\rmd C_{B1}\rmd\lambda_{B1}\rmd\gamma_{B1}\rmd\eta_{B1}\right)\\
 & \times\int\left(\rmd E_{B2}\rmd L_{B2}\rmd C_{B2}\rmd\lambda_{B2}\rmd\gamma_{B2}\rmd\eta_{B2}\right)\left(\rmd\ell_{b}\rmd g_{b}\right)\nonumber\\
 & \times\f{GM_{b}}{\left(2\mathcal{M}_{b}E_{b}\right)^{3/2}L_{b}}\f{GM_{B1}L_{B1}}{\left(-2\mathcal{M}_{B1}E_{B1}\right)^{3/2}}\f{GM_{B2}L_{B2}}{\left(-2\mathcal{M}_{B2}E_{B2}\right)^{3/2}}\nonumber
\end{align}
And then integrate over the $b$ angles, while, as before, limiting $\ell_{b}$ from $-\ell^{\left(b\right)}_{\max}$ to $\ell^{\left(b\right)}_{\max}$. We call that interval $2\D\ell_{b}$:
\begin{align}
\sigma & =M^{3}_{b}M^{3}_{B1}M^{3}_{B2}\int\left(\rmd E_{B1}\rmd L_{B1}\rmd C_{B1}\rmd\lambda_{B1}\rmd\gamma_{B1}\rmd\eta_{B1}\right)\nonumber\\
 & \times\int\left(\rmd E_{B2}\rmd L_{B2}\rmd C_{B2}\rmd\lambda_{B2}\rmd\gamma_{B2}\rmd\eta_{B2}\right)\nonumber\\
 & \times\f{GM_{b}}{\left(2\mathcal{M}_{b}E_{b}\right)^{3/2}L_{b}}\f{GM_{B1}L_{B1}}{\left(-2\mathcal{M}_{B1}E_{B1}\right)^{3/2}}\f{GM_{B2}L_{B2}}{\left(-2\mathcal{M}_{B2}E_{B2}\right)^{3/2}}\nonumber\\
 & \times\cdot2\pi\cdot2\D\ell_{b}\nonumber
\end{align}
At this point, the 3BP procedure is interrupted, since integrating over $\eta_{Bi}$ is not impossible. It seems innocuous, since as in the 2+1 case it appears here only inside $L_{b,x}^{2}+L_{y,b}^{2}$. But here it does not vanish:
\begin{align}
L^{2}_{b,x}+L^{2}_{y,b} & =\left(0-L_{B1,x}-L_{B2,x}\right)^{2}+\left(0-L_{B1,y}-L_{B2,y}\right)^{2}\nonumber\\
 & =\left(\sum_{Bi}L_{Bi}\s{1-C^{2}_{Bi}}\sin\eta_{Bi}\right)^{2}\nonumber\\
 & +\left(\sum_{Bi}L_{Bi}\s{1-C^{2}_{Bi}}\cos\eta_{Bi}\right)^{2}\nonumber\\
 & =L^{2}_{B1}\left(1-C^{2}_{B1}\right)+L^{2}_{B2}\left(1-C^{2}_{B2}\right)\nonumber\\
 & +2L_{B1}L_{B2}\s{\left(1-C^{2}_{B1}\right)\left(1-C^{2}_{B2}\right)}\cos\Delta\eta_{B}
\end{align}
The dependence on $\cos\left(\eta_{1}-\eta_{2}\right)$ prevents us from integrating over $\eta_{Bi}$ as we would like. We choose to just leave the integration there and perform it numerically when sampling from the distribution.

Another issue stems from the limits of the $\lambda_{Bi}$ angles. Identically to the 3BP formalism, these angles should be integrated from $0$ to $2\pi$. But doing so allows extremely soft or hard binaries into the calculation, even as these would never experience chaotic four-body interactions. This issue is expressed in an infrared divergence in the energies, where either $E_{Bi}\to 0$. To solve this issue, we put a limit on $\lambda_{Bi}$ in a similar manner to the limit on $\ell_{b}$. This new limit forces the components of such binaries to be close enough to experience full four-body chaos, but not so close that they will experience 3BP chaos instead. Mathematically speaking, it eliminates the divergence in the same way the $\ell_b$ eliminates the divergence in $E_b\to 0$.

The expression we get from these two modifications of the theory is:
\begin{align}
\sigma & =M^{3}_{b}M^{3}_{B1}M^{3}_{B2}\int\left(\rmd E_{B1}\rmd L_{B1}\rmd C_{B1}\rmd\eta_{B1}\right)\nonumber\\
 & \times\int\left(\rmd E_{B2}\rmd L_{B2}\rmd C_{B2}\rmd\eta_{B2}\right)\nonumber\\
 & \times\f{GM_{b}}{\left(2\mathcal{M}_{b}E_{b}\right)^{3/2}L_{b}}\f{GM_{B1}L_{B1}}{\left(-2\mathcal{M}_{B1}E_{B1}\right)^{3/2}}\f{GM_{B2}L_{B2}}{\left(-2\mathcal{M}_{B2}E_{B2}\right)^{3/2}}\nonumber\\
 & \times\left(2\pi\right)^{3}\cdot2\D\ell_{b}\cdot2\D\lambda_{B1}\cdot2\D\lambda_{B2}
\end{align}
Where $\D\lambda_{Bi}$ are the interval limits on $\lambda_{Bi}$, from $\lambda_{\min}$ to $\lambda_{\max}$. We simplify it and get:
\begin{align}
\sigma & =\int\rmd E_{B1}\rmd L_{B1}\rmd C_{B1}\rmd E_{B2}\rmd L_{B2}\rmd C_{B2}\nonumber\\
 & \times\left(\s 2\pi G\right)^{3}\f{M^{4}_{b}M^{4}_{B1}M^{4}_{B2}}{\mathcal{M}^{3/2}_{b}\mathcal{M}^{3/2}_{B1}\mathcal{M}^{3/2}_{B2}}\f{L_{B1}L_{B2}}{\left(E_{b}E_{B1}E_{B2}\right)^{3/2}}\nonumber\\
 & \times\f{\D\ell_{b}\D\lambda_{B1}\D\lambda_{B2}}{L_{b}}\rmd\eta_{B1}\rmd\eta_{B2}
\end{align}
The integrand is the distribution we are after. So we can finally write:
\begin{align}
 & \frac{\rmd\sigma}{\rmd E_{B1}\rmd L_{B1}\rmd C_{B1}\rmd E_{B2}\rmd L_{B2}\rmd C_{B2}}=\nonumber\\
 & \left(\s 2\pi G\right)^{3}\f{M^{4}_{B1}M^{4}_{B2}M^{4}_{b}}{\mathcal{M}^{3/2}_{B1}\mathcal{M}^{3/2}_{B2}\mathcal{M}^{3/2}_{b}}\f{L_{B1}L_{B2}}{\left(E_{B1}E_{B2}E_{b}\right)^{3/2}}\nonumber\\
 & \times\int\f{\D\lambda_{B1}\D\lambda_{B2}\D\ell_{b}}{L_{b}}\rmd\eta_{B1}\rmd\eta_{B2}
\end{align}
the observant reader may see the similarity to the 3BP case, in appendix \ref{app:21-derivation} (Eq. \ref{eq:21-general}). This is basically the same expression, with one more power of $\sqrt{2}\pi G$, more elements in each term, and more integrals. The $\pi$ that appears in the 3BP result is equivalent to $\Delta \lambda_B$, which is not needed there.

\bsp
\label{lastpage}

\end{document}